\documentstyle[12pt]{article}

\begin{document}

\markboth{E. M. Monte  \hspace{1mm}and\hspace{1mm}  M. D. Maia} 
{On Schwarzschild's Topology in Brane-Worlds}

\title{ON SCHWARZSCHILD'S TOPOLOGY IN BRANE-WORLDS}
\author{E. M. MONTE\thanks{e-mail: edmundo@fisica.ufpb.br}\\
Departamento de Fisica, Universidade \\Federal da Paraiba, 58059-970, Jo\~{a}o Pessoa, Paraiba, Brazil.\\
and\\
 M. D. MAIA\thanks{e-mail: maia@fis.unb.br}\\
Instituto  de Fisica, Universidade de Brasilia, 70919-970,
 Brasilia, \\Distrito Federal, Brazil.}

\maketitle

\begin{abstract}
The topological structure of Schwazschild's space-time and  its maximal analytic extension are investigated in the context of brane-worlds. Using the  embedding coordinates, these geometries are seen as  different states of the evolution of a single brane-world.
Comparing the  topologies and the embeddings  it is shown that this  evolution  must be followed by  a  signature change in the bulk.

\end{abstract}

\section{Introduction}	

The emphasis in the development of higher dimensional theories
has recently shifted towards  the brane-world picture\cite{ADD,RS}. A brane-world may be regarded as a space-time locally embedded in a higher dimensional space,  the bulk,  solution of higher dimensional Einstein's equations. Furthermore, the embedded geometry  is assumed to  exhibit quantum  fluctuations
with respect to  the extra dimensions at   the  TeV scale of  energies. Finally,  all  gauge interactions belonging to the standard model must remain confined to the four-dimensional space-time. Contrasting with  other  higher dimensional theories, the extra dimensions may be large and even infinite, with the possibility of being observed by  TeV accelerators.  

The embedding  conditions relate the  bulk  geometry  to the brane-world geometry, as it is clear from the  Gauss-Codazzi-Ricci  equations.  Therefore, it is likely that the quantum fluctuations of  the  embedded geometry  affect somehow the geometry of the bulk.  The purpose of this paper is to  show this relationship through a particular example, where we have an explicit embedding.

Since  a quantum theory of embedded geometries is not yet available, it is costmary to  work  in brane-worlds  with the classical approximation, in the form of geometric perturbations. 
In particular, for   a given embedded background geometry,  we may generate a  coordinate gauge independent perturbation of  that background by  a local  shift of the embedding  along the  extra  dimensions. The result   is another geometry representing the evolution of the original  brane-world\cite{ME}.

In our example  we take as the  background  a  spherically symmetric brane-world (Schwarzschild).  Then  we perturb  this  geometry by  extending it to  the   maximal  analytic extension represented by  Kruskal's  metric.  By comparing  the embedding and  topological differences  between the  background and the 
perturbed geometries,  we conclude that the perturbation induce a 
change the in signature of the bulk. 

Using the usual spherical coordinates $(t,r,\theta ,\phi)$ the  Schwarzschild space-time is $(E,g)=((V_{4}\cup B_{4}),g)$, with metric $g$,  

\begin{equation}
ds^{2}=(1-2m/r)dt^{2}-(1-2m/r)^{-1}dr^{2}-r^{2}(d\theta ^{2}+sin^{2}\theta
d\phi ^{2}). 
\end{equation}

Using its intrinsic properties, we may separate this  space into two regions:\cite{Oneill,MEtop}

i) The exterior Schwarzschild space-time $(V_{4},g)$:
\[
V_{4}=P_{I}^{2}\times S^{2}\;, \;P_{I}^{2}=\{(t,r)\in
I\!\!R^{2}|\;r>2m\}.
\]

ii) The Schwarzschild black hole $(B_{4},g)$:
\[
B_{4}=P_{II}^{2}\times S^{2}\;, \;P_{II}^{2}=\{(t,r)\in
I\!\!R^{2}|\;0<r<2m\}.
\]
In both  regions  $S^{2}$  denotes  the sphere of radius $r$. 

The  embeddings that are derived  from the solutions of the Gauss-Codazzi-Ricci equations, without use of  additional  conditions are  called differentiable embeddings. This is  the  kind of  embedding that is  appropriate  to  describe the local evolution  of brane-worlds, because an otherwise rigid or an  analytic embedding would be too  simple  or  too specific to generate   high frequency brane-worlds  fluctuations with  TeV  energy, including their transient conditions.

 Some relevant results  are  derived  from   differentiable  embedding techniques:   Collinson's theorem\cite{Collinson}  state that no non-flat empty space-time can be embedded locally and isometrically in a five-dimensional space of nonzero constant curvature. A theorem due to Szekeres tells  that  no non-flat vacuum metric can be embedded in five dimensions\cite{Szekeres}. 
 Therefore, the Schwarzschild's background cannot be differentiably embedded in a five dimensional flat or  constant curvature bulk, as in the  Randall-Sundrum model. For  simplicity,  we will consider the case of six-dimensional flat bulks $M_{r,s}$ with signatures $(r,s)=(4,2)$ and $(r,s)=(5,1)$,   where  the  Schwarzschild  space-time is known to be embeddable.

\section{The Embedding  Structure}

The two known local immersions of the Schwarzschild  space-time into  six  dimensional pseudo-Euclidean bulks are: 
(using $2m=1$) \newpage

%\vspace{2mm}  
The Kasner embedding \cite{Kasner} 
\[
Y:  (E,g)\rightarrow M_{4,2}\;\;
\left\{ 
\begin{array}{l}
Y_{1}=(1-1/r)^{1/2}\mbox{cos}t \\ 
\vspace{1mm}Y_{2}=(1-1/r)^{1/2}\mbox{sin}t \\ 
\vspace{1mm}Y_{3}=f(r),\;\;(df/dr)^{2}=\frac{1+4r^{3}}{4r^{3}(r-1)} \\ 
\vspace{1mm}Y_{4}=r\mbox{sin}\theta \mbox{sin}\phi  \\ 
\vspace{1mm}Y_{5}=r\mbox{sin}\theta \mbox{cos}\phi  \\ 
\vspace{1mm}Y_{6}=r\mbox{cos}\theta \vspace{1mm}
\end{array}
\right \}
\]
%\[
%ds^{2}=\;dY_{1}^{2}+dY_{2}^{2}-dY_{3}^{2}-dY_{4}^{2}-%dY_{5}^{2}-%dY_{6}^{2}.
%\]

and  the Fronsdal embedding \cite{Fronsdal}  
\[
Y^{\prime}:  (E,g)\rightarrow M_{5,1}\;\;
\left\{ 
\begin{array}{l}
Y_{1}^{\prime }=2(1-1/r)^{1/2}\mbox{sinh}(t/2) \\ 
\vspace{1mm}Y_{2}^{\prime }=2(1-1/r)^{1/2}\mbox{cos}h(t/2) \\ 
\vspace{1mm}Y_{3}^{\prime }=g(r),\;\;(dg/dr)^{2}=\frac{(r^{2}+r+1)}{r^{3}}
\\ 
\vspace{1mm}Y_{4}^{\prime }=r\mbox{sin}\theta \mbox{sin}\phi  \\ 
\vspace{1mm}Y_{5}^{\prime }=r\mbox{sin}\theta \mbox{cos}\phi  \\ 
\vspace{1mm}Y_{6}^{\prime }=r\mbox{cos}\theta \vspace{1mm}
\end{array}
\right \}.   
\]
%\[
%ds^{2}=\;{d{Y^{\prime }}_{1}}^{2}-{d{Y^{\prime }}_{2}}^{2}-{d%{Y^%{\prime %}}
%_%{3}}^{2}-{d{Y^{\prime }}_{4}}^{2}-{d{Y^{\prime }}_{5}}^{2}-%{d%%{Y^{\prime }}%
%_{6}}^{2},%
%\]
Notice that $Y_{3}^{\prime }$ is defined for $r>0$, while $Y_{3}$ is
defined only for $r>1$, suggesting the extension of $(E,g)=((V_{4}\cup B_{4}),g)$.

Now  we can show that  the  Fronsdal embedding   indeed  corresponds to an extension of the 
Schwarzschild  metric.  For this purpose,  define the new coordinates $u$ and $v$ by: For $r>2m$, 
\begin{equation}
v=\frac{1}{4m}(\frac{r}{2m})^{1/2}exp(\frac{r}{4m}){Y^{\prime }}_{1}\;\;%
\mbox{and}\;\;u=\frac{1}{4m}(\frac{r}{2m})^{1/2}exp(\frac{r}{4m}){Y^{\prime }%
}_{2}.
\end{equation}
For $0<r<2m$, 
\begin{equation}
v=\frac{i}{4m}(\frac{r}{2m})^{1/2}exp(\frac{r}{4m}){Y^{\prime }}_{1}\;\;%
\mbox{and}\;\;u=\frac{i}{4m}(\frac{r}{2m})^{1/2}exp(\frac{r}{4m}){Y^{\prime }%
}_{2},
\end{equation}
where 
\begin{equation}
u^{2}-v^{2}=(\frac{r}{2m}-1)exp(\frac{r}{2m})\;\Longleftrightarrow \;{%
Y^{\prime }}_{2}^{2}-{Y^{\prime }}_{1}^{2}=16m^{2}(1-\frac{2m}{r}).
\end{equation}
Now $r=r({Y^{\prime }}_{1},{Y^{\prime }}_{2})$ is implicitly defined by
equation $(4)$, while $t=t({Y^{\prime }}_{1},{Y^{\prime }}_{2})$ is
implicitly defined by 
\begin{equation}
{Y^{\prime }}_{1}/{Y^{\prime }}_{2}=tgh(\frac{t}{4m}).\label{eq:tgh}
\end{equation}
Finally, the metric $g^{\prime }$ in the new coordinates is 
\begin{equation}
ds^{2}=(32m^{3}/r)exp(-r/2m)(dv^{2}-du^{2})-r^{2}(d\theta ^{2}+sin^{2}\theta
d\phi ^{2}),
\end{equation}
which is  the same metric encountered by Kruskal\cite{Kruskal}  representing the maximall
analytic extension of the Schwarzschild metric without a singularity at $r=2m$. Using the new coordinates $(u,v,\theta ,\phi)$, the Schwarzschild space-time $(E,g)$ is extended to space-time $(E^{\prime }=Q^{2}\times S^{2},g^{\prime})$, where $Q^{2}$ is the Kruskal plane  defined by $(4)$.

\section{The Topological Structure}

In the following we show that the topology of $(E,g)$ is $%
(\{I\!\!R^{2}-\{(t,r)\in I\!\!R^{2}|\;r=2m\})\times S^{2}$ and that the topology
of  $(E^{\prime },g^{\prime })$ is $I\!\!R^{2}\times S^{2}$, when the space $(E,g)$ is embedded in differents bulks. Indeed, by construction: $E=[P_{I}^{2}\cup P_{II}^{2}]\times S^{2}$ and $E^{\prime }=Q^{2}\times S^{2}$. The topology of $E$ is the Cartesian product
topology of $[P_{I}^{2}\cup P_{II}^{2}]$ by $S^{2}$, while that the topology
of $E^{\prime }$ is the Cartesian product topology of $Q^{2}$ by $S^{2}$. While the topology  $S^{2}\subset I\!\!R^{3}$ is the usual topology\cite{Dugundji} induced by the topological space $(\tau _{3},I\!\!R^{3})$, 
 the topologies of $[P_{I}^{2}\cup P_{II}^{2}]\subset I\!\!R^{2}$ and
of $Q^{2}\subset I\!\!R^{2}$, respectively, $\tau _{p}$ and $\tau _{q}$, are induced from $(\tau _{2},I\!\!R^{2})$. 

Since $Q^{2}$ is an extension of $[P_{I}^{2}\cup P_{II}^{2}]$, we may define the  embedding,  $\psi :[P_{I}^{2}\cup P_{II}^{2}]\;\rightarrow \;Q^{2}$.
Therefore,  defining  the  open set $A=\{(t,r)\in I\!\!R^{2}|\;t^{2}+(r-2m)^{2}<m^{2}\;\,\mbox{and}\;\,r>0\}$, we have $A\cap \lbrack P_{I}^{2}\cup P_{II}^{2}]=A-\{(t,r)\in
I\!\!R^{2}|\;r=2m\}$. This is an open set of the topological space $
[P_{I}^{2}\cup P_{II}^{2}]$, composed of two connected components  which 
form a topological basis for the semi-plane $t-r,\;r>0$.

On the other hand, we have that $\psi (A\cap \lbrack P_{I}^{2}\cup P_{II}^{2}])$ is
given for another  open set composed by four connected components. Using $(4)$ it follows that  the lines $L_{1}$ and $L_{2}$ defined corresponding to  $r=2m$ are on $Q^{2}$, we have that $\{[\psi (A\cap \lbrack P_{I}^{2}\cup P_{II}^{2}])\cup L_{1}\cup L_{2}]\}\cap
D=B$, where $D$ is an open disk on $I\!\!R^{2}$ with center in the origin of $Q^{2}$. Therefore,  $B$ is a plane disk  in the new coordinates $r=r({
Y^{\prime }}_{1},{Y^{\prime }}_{2})$ and $t=t({Y^{\prime }}_{1},{Y^{\prime }}%
_{2})$. Consequently,  the topology of $Q^{2}$ is given by open sets of $
I\!\!R^{2}$.

Our  conclusion is that  the topology of  $(E,g)$, $
(I\!\!R^{2}-\{(t,r)\in I\!\!R^{2}|\;r=2m\})\times S^{2}$, is
different from the topology $I\!\!R^{2}\times S^{2}$  of the extension $(E',g')$. Notice that $(E,g)$ is composed by two components and consequently it is disconnected, while $(E^{\prime},g^{\prime })$ is connected as it follows from the Fronsdal embedding.

While the embedding   of  the Schwarzschild  space-time given by Kasner  remains differentiable only   for  $r>2m$,  or in another  connected component $0<r<2m$, the  Fronsdal  embedding  remains differentiable all the way to $r=0$ (for $r>0$ in a only connected component). If the latter is regarded as  an  evolution of the  former, then   this is  only possible  with    a change of  signature  and topology of bulk.  These changes should
be observable at  the predicted laboratory production of  black-holes.

\end{document}